\documentstyle[12pt]{article}
\begin{document}
\pagestyle{empty}
\begin{flushright}
{CERN-TH.6719/92} \\
\end{flushright}
\vspace*{5mm}
\begin{center}
{\bf White Noise and Heating of Open Quantum Field Fermi Systems} \\
[1 0mm]
A.A.Abrikosov, Jr.  \footnote{Permanent address: Institute of Theoretical
and Experimental Physics, Bolshaya Cheremushkinskaya ul. 25, 117259, Moscow,
Russia. \\E-mail: {\small PERSIK@VXITEP.ITEP.MSK.SU; ABRIKO@VXCERN.CERN.CH}} \\
[10mm]
Theory Division\\
CERN, CH-1211 Geneva 23, Switzerland \\[2cm]
{\bf Abstract}
\end{center}
\vspace*{3mm}
Stochastic external fields could represent the influence of a heat
bath on quantum statistical systems. We study the time evolution of the
density matrices of quantum Fermi systems interacting with classic external
Fermi fields. This interaction can either change the temperature of
the system or it can affect the density of
particles. In the case of relativistic Dirac fermions, variations of
temperature lead to creation (annihilation) of particle - antiparticle
pairs. The change of the density (or of the chemical potential)
 indicates the existence of the
incoming (outgoing) flux of fermions from (to) the bath.
These changes are independent for the different modes and in order
 to model the
thermalization one should adjust the spectrum of the noise.  The linear
time dependences of the densities of particles are characteristic for
all the processes. \\[2cm]
\begin{flushleft}
CERN-TH.6719/92 \\
November 1992 \\
\end{flushleft}
\thispagestyle{empty}\mbox{}

\newcommand{\hh}{\hat{H}}
\newcommand {\tr}{\,{\rm tr}\,}
\newcommand{\Tr}{\,{\rm Tr}\,}
\newcommand {\p} {\psi}
\newcommand{\bb}{\begin{equation}}
\newcommand{\ee}{\end{equation}}
\newcommand {\arctanh} {\,{\rm arctanh}\,}
\def\atlim#1#2
{\hskip.4ex \hbox{$
\raise-2.2ex\hbox{\vrule height 5.6ex width.05ex}
\hskip-.4ex
\raise 3ex\hbox{
$\scriptstyle#1$}\hskip-0.8ex
\raise-2.8ex\hbox{$\scriptstyle#2$}
$}}

\newcounter{ap}{}
\newcounter{aq}{}
\newcounter{sap}{}
\newcommand{\saskip}{}

\newcommand{\app}[1]{\addtocounter{ap}{1}
\vspace{5mm}
\begin{center}
{\bf Appendix {\Alph{ap}}}\vspace{3mm}\\
{\bf {#1}}\vspace{2mm}
\nopagebreak
\end{center}
\setcounter{aq}{0}
\renewcommand{\saskip}{\vspace{-6mm}}
\renewcommand{\theequation}{\addtocounter{aq}{1}\Alph{ap}.\theaq}}

\newcommand{\subapp}[1]{\addtocounter{sap}{1}
\saskip
\begin{center}
{\bf \alph{sap})} \hspace{2pt} {\bf {#1}}\vspace{1mm}
\nopagebreak
\end{center}
\renewcommand{\saskip}{\vspace{4mm}}}

\newpage\setcounter{page}{1}\pagestyle{plain}
\section{Introduction}

Interest in the problems of thermalization and heat transfer in
statistical systems is due to a number of reasons. Not to mention
natural scientific curiosity, there are the following more pragmatic
circumstances.

The first comes from the wide application of numerical methods in
quantum field theory and statistical physics. The greater part of the
methods study systems contacting the heat bath. The efficiency
of these calculations strongly depends on the model of the interaction
of the system with the environment. Thus the search for better models
of heat exchange is an important way to improve these numerical
approaches.

Secondly, the investigation of systems with varying temperature can
be a bridge to the domain of strong fluctuations in the vicinity
of second-kind phase transitions alternative to the fixed temperature
approach. The idea is that by introducing external heat sources one can scan
the dangerous region in finite time. By adjusting the power of the sources
this can be done fast enough so the growth of fluctuations does not spoil
the validity of the approximations used.

Certainly all the thermalization models should include assumptions
about some irreversible processes since the time-reversibility of
quantum mechanics forbids us to build them from first principles.

The present paper studies the possibility to represent the influence
of a heat bath by means of a stochastic external field. The Bose-case
was investigated in the paper \cite{aaa} for examples of the harmonic
oscillator and of the free bosonic field. The presence of white
noise immediately reminds us of stochastic quantization \cite{bb}. However,
dissipation is now absent and exponential relaxation is replaced
by continuous energy pumping. The number of quanta per degree of
freedom grows linearly and heating to the desired
temperature takes a finite time. The
apparent flaw is that different modes are heated independently and
one has to adjust the spectrum of the sources in order to imitate
thermalization.

Now we extend the discussion to the free fermionic systems interacting
with stochastic Fermi fields. We begin from one degree of freedom and
then generalize the analysis to Dirac fermions. The results are
greatly analogous to those listed above. In the simplest case the external
source causes the linear increase (or decrease for the noise constant
of the opposite sign) of the fermion number. In the
second case one can split the noise into the two components. The
first creates fermion-antifermion pairs while the second generates the
particle-antiparticle asymmetry. The densities of both particles and
antiparticles change linearly just as in the first case.

The serious specifics of the Fermi case come from the Pauli
principle. There is nothing unusual in the linearly growing number of
bosons per state. However, the average occupation number of fermions
should not exceed one. If the latter change linearly then the boundaries
are attained in finite time. That means that the model has limited
virtues even in the simplest case of one degree of freedom.

As an improvement it seems natural to include the backward action of the
system on the heat bath. For example, in the procedure of stochastic
quantization the dissipation constant is introduced for this purpose.
A less phenomenological alternative is to link the intensity of the noise
with occupation numbers. However, it may be more reasonable to specify the
amendments bearing in mind only the particular applications.

The paper has the following structure. We recall some basic definitions and
properties of the thermodynamical density matrix in
Section 2. Section 3 deals with the problem for a single
fermion and Section 4 describes
the heating of the Dirac field, firstly in general and
then in more detail. Some
formulae used in the  text are reviewed in the Appendices which follow the
concluding Sections.

\pagestyle{plain}
\section{The Density Matrix in the Presence of the Stochastic External Field.}

We shall study how the statistical density matrix changes under the action
of the random classical fields. We shall use the technique of the
influence functional by Feynman and Vernon \cite{aaa},\cite{cc}, the essence of
which we shall briefly recall now. Let us begin from the definitions.

The wave functions $\p$ of the quantum system with the Hamiltonian $\hh$
satisfy the Schr\"{o}dinger equation:
\bb
\hh \p _i=E_i \p _i
\ee
where $i$ is a level number and $E_i$ is its energy.

The expansion of the equilibrium thermal density matrix in the basis of the
eigenfunctions $\p _i$ contains only diagonal terms \cite{dd}:
\bb
\rho_{\beta}(x, x^{\prime})\propto \sum^{\infty}_{i=1}\psi_i(x)\psi^{\ast}_i
(x^{\prime}) \exp(-\beta E_i)
\ee
Here $\beta =\frac{1}{T}$ is the inverse temperature and the summation is
over all states. If we normalize the density matrix so that $\tr
\rho_{\beta}=1$,
then the diagonal terms are the statistical probabilities of states. (Note that
for fermions the fermionic number operator $\hat{F}$ should be included in the
normalization condition which  becomes $\tr (-1)^{\hat F} \rho_{\beta}=1$).

Actually the density matrix defined by Eq. 2 has the sense of the
average over the statistical ensemble, {\em i.e.} over a large number $n$
of identical systems immersed in the heat bath:
\bb
\rho_{\beta}(x, x^\prime)=\langle \rho (x, x^\prime) \rangle =
\lim_{n\rightarrow \infty}\frac{1}{n}\sum^{n}_{k=1} \rho_k(x, x^\prime)
\ee

Another way to define $\rho_{\beta}$ is to average the thermally
fluctuating density matrix over a sufficiently long period. But the
definition (3) is more  convenient for the study of the time evolution.

The objects of quantum kinetics are density matrices $\rho (x, t; x^\prime,
t^\prime) $ depending on two time arguments \cite{ee}. Each of the $\rho _k
(x, t; x^\prime, t^\prime) $ matrices satisfies the system of equations:
\bb
\begin{array}{rccccr}
i\hbar\frac{\partial}{\partial  t}\rho_k(t, t^{\prime}) & = & i \hbar
\stackrel{\rightarrow}{\partial} \rho_{k} (t, t^{\prime}) & = &
\hat{H}\rho_k(t,t^{\prime}) & ~~~~~~~~~~~~~~~~~~~~~~ (a) \\
-i\hbar\frac{\partial}{\partial t^{\prime}}\rho_k(t,t^{\prime}) & = & i \hbar
\rho_{k} (t, t^{\prime}) \stackrel{\leftarrow}{\partial} & = &
\rho_k(t,t^{\prime})\hat{H} & ~~~~~~~~~~~~~~~~~~~~~~~(b)
\end{array}
\ee

In an open system the Hamiltonian is a function of time and $\rho_k$
depends on both arguments $t$ and $t^\prime$ (for $\hh = \hh_{0} = const$ it
would
in fact be a function only of the difference $t-t^\prime$).
The general solutions of the above equations
are well known and can be expressed in terms of the evolution operator
$\hat{S}$
\cite{ff},\cite{ggg}:
\bb
\hat{S}(t,t^{\prime})= \exp[ - i\int\limits^{t}_{t^{\prime}}
\hat{H}(\tau)d \tau]
\ee
Namely $\p(t)=\hat{S}(t, t_0) \, \p(t_0)$,  $\p^\ast (t) = \p^\ast (t_0) \,
\hat{S}^\dagger
(t_0, t)$. Here $\hat{S}^\dagger$ denotes the Hermitian conjugated
operator. In calculations we shall use the integral representation of the
operator $\hat{S}$.
For Bose systems, the kernel $S$ can be represented as a path integral along
the trajectories connecting $x$ and $x^\prime$.
\bb
S(x, t; x^{\prime},t^{\prime})=\int [\frac{dp \, dq}{2\pi}] \exp \{i\int\limits
^{t}_{t^{\prime}}[p \dot{q}-H(p,q)]d \tau \}
\ee
where $p$ and $q$ stand for generalized  momenta and coordinates.

For Fermi-systems the amplitude of the transition from the initial state
$\p(t^\prime)=\p
_i$ to the final state $\p^\ast (t)=\p^\ast  _f$ is given by the integral over
Grassmann variables
\bb
S(\psi^{\ast}_f, t; \psi_i, t^{\prime})=\int[d \psi^{\ast} d\psi]
\exp\{\psi^{\ast}(t)\psi(t)+i\int\limits^t  _{t^{\prime}}
[i\psi^{\ast}(\tau)\dot{\psi}(\tau) -H(\psi^{\ast}, \psi)]d \tau \} \ee

The solution of the system of eqs.(4) for the Bose case is
\bb
\rho_k(t,t^{\prime})=\hat{S}(t,t_0) \,\rho(t_0) \, \hat{S}^{\dagger} (t_0,
t^{\prime})
\ee

The operator $\hat{S}$ is unitary by definition, $\hat{S}\,\hat{S}^\dagger =1$
and all the matrices $\rho_k$
remain normalized as time elapses:
\bb
\tr \rho_k(t)=\tr \rho_k(t_0)=1
\ee

This means that the trace of the average $\rho_\beta$ is constant too.

The evolution of density matrices of Fermi systems looks similar,
but for the
factor $(-1)^{\hat F}$ . To show how it appears let us return to the definition
of
$\rho $.

The ensemble average of some quantity $\langle A  \rangle $ is given by the
formula
(let us assume that we start from the diagonal $\rho$-matrix)
\bb
\langle A  \rangle = \sum_{i=0}^\infty W_i <\psi^{\ast}_i\mid \hat{A} \mid
\psi_i >
=\tr(-1)^{\hat F} \hat{A} \rho \ee
Here $\hat {A}$ is the quantum mechanical operator and $W_i$ are the
statistical probabilities which are proportional to $\exp (-\beta E_i)$
in equilibrium. The anticommuting $\p$ and $\p^\ast $ give  the $(-1)^{\hat F}$
mentioned above.

If the external field can produce the fermions then the evolution operator
$\hat{S}$ does not commute with $(-1)^{\hat F}$ and the location of the latter
is
significant,
\bb
(-1)^{\hat F} \hat{S}(t_1, t_2)\neq \hat{S}(t_1, t_2)(-1)^{\hat F}
\ee

Recalling the evolution laws of the  wave functions we write
\begin{eqnarray}
\lefteqn {\langle A (t) \rangle   =   \sum_{i=0}^\infty W_i (t_0)
< \psi^{\ast}_i(t_0) \hat{S}^{\dagger}(t_0,t) \mid \hat{A} \mid
\hat{S}(t,t_0)\psi_i (t_0)> =}
 & &\\
 & \hspace{5mm}= & \sum _{i=0}^{\infty}  W_i(t_0) \,  \tr [ \: \hat{A} \mid
\hat{S}(t,t_0)
\psi_i (t_0)> (-1)^{\hat F} \! <\psi^{\ast}_i(t_0) \hat{S}^{\dagger} (t_0,t)
\mid
\,] \nonumber
\end{eqnarray}

Now it is clear that if the fermion number is not conserved one should study
the behaviour of the density matrix $\tilde \rho (\p^{\ast}, \p; t)=(-1)^{\hat
F}
\rho  (\p^{\ast}, \p; t)$. The evolution law has the form
\bb
\tilde{\rho}(t) =\hat{S}(t,t_0) \, \tilde{\rho}(t_0) \,
\hat{S}^{\dagger}(t_0,t)
\ee
and  preserves the normalization of the latter: $\tr \tilde{\rho}(t)=const$.
\footnote{This definition is good enough for our problem since thermal
$\tilde{\rho}$ matrices do not have off-diagonal elements. Otherwise
$(-1)^{\hat F} \rho \not= \rho (-1)^{\hat F}$ could be the case.  Then it seems
impossible to express $\tilde{\rho}$ in terms of $\rho$.}

When passing from the
quantum mechanical matrices $\rho_k$ to  thermodynamical $\rho_\beta$ the
problem of
averaging over stochastic  fields arises. The Hamiltonian of the $k$-th system
is \bb
\hat{H}_k (t)=\hat{H}_0 + \hat{V}_k (t)
\ee
where $\hat{V}_k$ stands for the particular realization of the external
influence.
According to the general principles the following average should be calculated:
\bb
\langle \rho(t) \rangle =\lim_{n\rightarrow\infty} \frac{1}{n} \sum _{k=0}^n
\hat{S}_k (t,t_0) \,\rho_k(t_0)
 \, \hat{S}^{\dagger}_k (t_0,t)
\ee
for the thermally equilibrium initial state, {\em i.e.} $\rho(t_0)  =
\rho_\beta$ .

The question is whether the operators $\hat{S}_k$ are independent of  the
matrices
$\rho_k$ or whether the backward influence of the system on the heat bath is
important.

We shall focus on the simple model of the exterior which is absolutely ignorant
 of
the object. In this approximation fluctuations of $\hat{V}_k$ are independent
of
$\rho _k$. That gives:
\bb
\rho(t)=\langle \hat{S}(t,t_0) \, \rho_{\beta} \, \hat{S}^{\dagger}(t_0,t)
\rangle.
\ee
The averaging of the direct product  $\langle \hat{S}(t,t_0) \otimes
\hat{S}^{\dagger} (t_0,t) \rangle$ reflects the classical  origin
of the  field of the heat bath (in the quantum case the averages of $\hat{S}$
and
$\hat{S}^\dagger$ would be computed separately).

It was shown that in the bosonic case noise steadily pumps energy into the
system
and it is heated \cite{aaa}. That means that the classical bosonic noise always
models the environment of higher temperature. It is not necessarily so if both
the
system and the noise are the Fermi fields.

\section{The Interaction of Fermions With the Fermionic White Noise. One Degree
of Freedom.}

The case of one fermionic degree of freedom analysed here is readily
generalized to free fermionic systems.  We shall show that classical
external Fermi noise affects the density matrix and changes the effective
temperature.

There are only two vectors of state and the Hamiltonian written in terms of
creation and annihilation operators ${\hat\p}^{+}$ and ${\hat \p}$ is the
simplest one:
\bb
\hat{H}=m \hat{\psi}^{+}\hat{\psi} + \hat{V}(t)
\ee
Here $m$ is the energy of the excitation.

We shall use the holomorphic representation (see Appendix A) where the
Grassmann binomial
 $f (\p^{*}) = f^{0} + \p^{*}f^{1} $ corresponds to the mixed state
$| f >  = f^{0} |0 > + f^{1} |1 >$ \cite{ggg}.  The Hamiltonian is
the Grassmann integral operator with the kernel
\bb
H(\psi^{\ast}, \psi)=m \psi^{\ast}\psi + V(\psi^{\ast}, \psi; t)
\ee
The thermodynamical density matrix is
\bb
\rho_{\beta}(\psi^{\ast},\psi)=\frac{1}{1+e^{-\beta m}}(1+\psi^{\ast}
e^{-\beta m} \psi)=\frac{\exp \psi^{\ast}e^{-\beta m}\psi}{1+e^{-\beta m}}
\ee
and the operator $(-1)^{\hat F}$ has the form
\bb
(-1)^{\hat F} (\psi^{\ast}, \psi)= 1 - \psi^{\ast}\psi = \exp -\psi^{\ast}
\psi   \ee
We shall represent the outer influence by the following addition to the
Hamiltonian:
\bb
V(\psi^{\ast}, \psi; t) = \sigma^{\ast}(t) \psi + \psi^{\ast} \sigma (t)
\ee
The fields $\sigma (t), \sigma^{*}(t)$ are characterized by
their correlation function.  In the case of the fully
stochastic ({\em i.e.}  white noise) fields, it has the form
\bb
\langle \sigma^{\ast}(t) \sigma(t^{\prime})\rangle = \epsilon
\delta(t-t^{\prime})
\ee
The value of $\epsilon$ measures the power input and, as will be seen
later, fixes the time scale, while $m$ proves to be the energy and
temperature scale. The evolution operator for the transition from
$\p_{0}$ to $\p_{t}$ in time $\epsilon t$ when the noise is present is
(see Appendix B)
\begin{eqnarray}
\lefteqn {S(\psi^{\ast}_t,t; \psi_0,0)= \exp\{\psi^{\ast}_t e^{-imt} \psi_0 } &
& \\ & \hspace{5mm}  +i \int \limits^{t}_0
[\sigma^{\ast}{\bar e}^{im\tau}\psi_0+\psi^{\ast}_t
{\bar e}^{im(t-\tau)}\sigma] d\tau
 -\int \limits^{t}_0 d\tau_1 \int \limits^{\tau_1}_0
\sigma^{\ast}(\tau_1) e^{-im(\tau_1-\tau_2)}
\sigma(\tau_2)d\tau_2\} & \nonumber
\end{eqnarray}
and the Hermitian-conjugated one equals
\begin{eqnarray}
\lefteqn{S^{\dagger}(\psi^{\ast}_0,0; \psi_t,t)= \exp\{\psi^{\ast}_0
e^{imt}\psi_t
} & & \\
& \hspace{5mm} -
i\int\limits^t_0 [\sigma^{\ast}e^{im(t-\tau)}\psi_t +\psi^{\ast}_0 e^{im\tau}
\sigma]d\tau -
\int\limits^{t}_0 d\tau_1 \int\limits^t _{\tau_1} \sigma^{\ast}(\tau_1)
e^{-im(\tau_1-\tau_2)}\sigma(\tau_2) d\tau_2 \} & \nonumber
\end{eqnarray}
According to the previous section, we shall observe the changes of the
matrix $\tilde{\rho} = (-1)^{\hat F} \rho$ starting from the equilibrium
meaning
$\tilde{\rho}_{\beta}$ given by Eq. (19). Substituting into Eq. (13) the
expressions (23, 24) for $S, S^{\dagger}$ and using the rules of Appendix A, we
find the convolution
\begin{eqnarray}
\lefteqn{\tilde{\rho}(\psi^{\ast},\psi;t)=} \nonumber  \\
&  \hspace{5mm} \lefteqn {\frac{1}{1+e^{-\beta m}}\exp
[-\psi^{\ast}e^{-\beta m}\psi+i \int \limits
^{t}_{0}\psi^{\ast}(1+e^{-\beta m})  e^{im(\tau-t)}
\sigma(\tau)d\tau -}  &   \\
 & & -i \int\limits^{t}_{0}\sigma^{\ast}(\tau)e^{-im(\tau-t)}(1+e^{-\beta m})
\psi d\tau
-\int\limits^{t}_{0} \int \limits^{t}_{0} d\tau_1 d\tau_2 \sigma^{\ast}(\tau_1)
e^{-im(\tau_1-\tau_2)}(1+e^{-\beta m})\sigma(\tau_2)] \nonumber
\end{eqnarray}
One can carry out the fluctuation averaging by means of the Gaussian path
integral
\bb
\langle \tilde{\rho} \rangle =\frac{\int[d\sigma^{\ast} d\sigma] \,
\hat{S} \,\tilde{\rho} \, \hat{S}^{\dagger}  \,
\exp({\epsilon}^{-1}\int\limits^{t}_{0}\sigma^{\ast}\sigma dt)}
{\int[d\sigma^{\ast}d\sigma] \exp({\epsilon}^{-1}\int\limits^{t}_{0}
\sigma^{\ast}\sigma dt)}
\ee
which generates the $\delta$-correlated $\sigma,
\sigma^{*}$ fields.  It proves convenient to use the Fourier expansion of
$\sigma^{*}$ and $\sigma$
\bb
\sigma(\tau)=t^{-1/2}\sum^{\infty}_{n=-\infty}
\sigma_n e^{-i (\frac{2\pi n}{t} +m)\tau};~~~
\sigma^{\ast}(\tau)= t^{-1 /2}\sum^{\infty}_{n=-\infty}
\sigma^{\ast}_n e^{i( \frac{2\pi n}{t}+ m)\tau}
\ee
The reason for doing so is that only the zero Fourier-components of the
fields do contribute, giving
\bb
\langle \tilde{\rho}(t) \rangle=\frac{1-\epsilon t(1+e^{-\beta m})}{1+e^{-\beta
m}}
\exp[- \psi^{\ast} \frac{e^{-\beta m}+\epsilon t (1+e^{-\beta m})}
{1-\epsilon t (1+e^{-\beta m})}\psi]
\ee
One can immediately prove that it is properly normalized.

Let us discuss the expression (28).  The first optimistic observation is that
the $\tilde{\rho}$ matrix remains diagonal.  That means that it can still be
parametrized by the single parameter $\beta^{\prime }$
\bb
e^{-\beta^{\prime }m}=\frac{e^{-\beta m}+\epsilon t(1+e^{-\beta m})}
{1-\epsilon t (1+e^{-\beta m})}
\ee

Solving this equation gives the value
\bb
\beta^{\prime}=\frac{2}{m}\arctanh (\tanh \frac{\beta m}{2}-2\epsilon t)
\ee
but the result is more handy if expressed in terms of the
upper-level occupation number, {\em i.e.}, the overall average number of
fermions
\bb
n(\beta) =\frac{1}{e^{\beta m}+1}
\ee
This quantity depends linearly on time
\bb
n(\beta) \atlim{t}{0} = \epsilon t
\ee
giving evidence for the constant rate of energy pumping into the system
$$\frac{dE}{dt}=m\epsilon.$$
We conclude that the external source is directly coupled to the
fermionic number and $\epsilon$ measures the intensity.  For positive
values of $\epsilon$, the number of particles grows, otherwise the
fermionic drain takes place.

The second fact following from Eq. (32) is that the model is applicable
only  for time intervals $t << |\epsilon |^{-1}$. That is just the
consequence of the inequality $\frac{1}{2} > \langle n(\beta, \mu)\rangle >
0$ coming from Fermi statistics. (The inclusion of the so-called ``negative
temperatures" for inverted occupancies $\frac{1}{2}~<~n(\beta, \mu)~<~1$
does not affect the reasoning.)

In the case of the linear law (32), the limiting values (the upper or the
lower depending on the sign of $\epsilon$) are attained in final time
$\tau \sim \epsilon^{-1}$.  This indicates the breakdown of the
phenomenological assumption that the influence of the heat bath is of
permanent intensity. We shall discuss the possible improvements
after describing the Dirac case.

\section{The Dirac Fermions.}
\subsection{General relations.}

We shall now analyse the effect of stochastic fields on the density
matrix of free Dirac fermions. We shall study the Lagrangian
\bb
L = \bar{\psi}(\hat{p}-m)\psi + \bar{\sigma}\psi + \bar{\psi}\sigma
\ee
where
$$
\hat{p} = \hat{p}_{\mu}\gamma^{\mu} =
i\gamma^{\mu}\partial_{\mu},~~
g^{\mu \nu} = (1, -1, -1, -1).
$$
Let us recall that $\bar{\p}=\p^{\dagger}\gamma_0$,
$\gamma^{\dagger}_0=\gamma_0$,  $\gamma^{\dagger}_{i}=-\gamma_{i}$.
The free Dirac Hamiltonian is
\bb
\hat{H}=\gamma_0(\gamma_i \hat{p}_i + m)
\ee
and its integral representation has the following Hermitian
Grassmannian kernel:
\bb
H(\bar{\psi}\gamma_0, \psi)=\bar{\psi}(\hat{p}_i \gamma_i + m)\psi
\ee
We perform as usual the Fourier expansion
\bb
\psi(x,t)=\sum_{\bf p} \psi_{\bf p}(t)e^{i p_i x_i};~~~
\bar{\psi}(x,t)=\sum_{\bf p} \bar{\psi}_{\bf p}(t)e^{-i p_i x_i}
\ee
The operator $\hat{p}_{i}$ acting on a plane wave gives a number
$ p_{i}$ and the Hamiltonian converts into the matrix $\hh_{\bf p}$:
\bb
\hat{H}_{\bf p}=\gamma_0(p_i \gamma_i + m)
\ee
The energy of a particle with momentum $\bf{p}$ equals $E_{\bf p} =\sqrt{{\bf
p}^2 +m^2}$. The external fields $\bar{\sigma}, \sigma$ are now also bispinors
and their correlation function is a matrix:
\bb
\langle \bar{\sigma}(t) \sigma(t^{\prime}) \rangle =
- \hat{\epsilon} \delta (t-t^{\prime})
\ee
It is easy to notice that
the modes $\p_{\bf p}, \bar{\p}_{\bf p}$ interact only with the components
$\bar{\sigma}_{\bf p}, \sigma_{\bf p}$ of the noise respectively.

Unlike in the previous section, the thermal density matrix of Dirac
fermions depends on two parameters, {\em i.e.}, on the inverse
temperature $\rho$ and the chemical potential $\mu$.  The latter is
the variable conjugated to the fermion number, $F$, which is defined as
the difference between the numbers of particles and antiparticles.
The fermion-antifermion pair creation does not change $\mu$. Up to
normalization, the equilibrium statistical density matrix is
\bb
\rho (\beta, \mu) \propto \exp [- \beta (\hh - \mu \hat{F}) ]
\ee
The corresponding kernel factorizes into the product over the modes
\bb
\rho(\bar{\psi}    \gamma_0,    \psi; \beta,     \mu)=\prod
\limits_{\bf p}  \rho_{\bf  p}(\beta,\mu)
=\prod\limits_{\bf p} \frac{\exp
\bar{\psi}_{\bf p}\gamma_0     e^{-\beta(\hat{H}_{\bf  p}-\mu)}\psi_{\bf p}}
{4e^{2\beta\mu}(\cosh \beta E +\cosh\beta\mu)^2}
\ee
The matrix exponent is easily obtained by means of the identity
$\hh^{2}_{\bf p} = E^{2}_{\bf p}$, giving for real and imaginary arguments
\bb
\begin{array}{lclcrr}
\exp(-\beta \hat{H}_{\bf p}) & = & \cosh\beta E_{\bf p} & - &
\frac{\hat{H}_{\bf p}}{E_{\bf p}}\sinh \beta E_{\bf p} & \hspace{10mm} (a)
\\
\exp(i \hat{H}_{\bf p} t) & = & \cos E_{\bf p}t & + & i\frac{\hat{H}_{\bf p}}
{E_{\bf p}} \sin E_{\bf p}t  & \hspace{10mm}(b)
\end{array}
\ee
The normalisation factor is obtained by the transformation
$$\det \left[\exp \beta (\mu - \hat{H}_{\bf p}) \right] =
[2 e^{\beta \mu} (\cosh \beta \mu + \cosh \beta E_{\bf p})]^{2}.$$

Actually, when working with the Hamiltonian we always use the complex
conjugate functions $\p^{\dagger} = \bar{\p} \gamma_{0}$ instead of
$\bar{\p}$ common for relativistic formulae. This follows from the
non-relativistic nature of the thermodynamics, where the reference
frame is always present.  However, the choice (39) enables us to
convolute $\rho$ directly with $S, S^{\dagger}$.

The calculation of the operator of evolution between the states $\p (0)
= \p_{0}$ and $\bar{\p} (t) = \bar{\p}_{t}$ is done in Appendix B.
\begin{eqnarray}
\lefteqn{S(\bar{\psi}_t \gamma_0, t; \psi_0,0)=\exp \{\bar{\psi}_t
\gamma_0  e^{-i\hat{H}t}\psi_0 } & & \\
 & \hspace{5mm} + i\int\limits^{t}_{0}
[\bar{\sigma}e^{-i\hat{H}\tau}\psi_0+\bar{\psi}_t \gamma_0
e^{-i\hat{H}(t-\tau)}\gamma_0 \sigma]d\tau -
 \int\limits^{t}_{0} d\tau_1 \int\limits^{\tau_1}_{0}\bar{\sigma}(\tau_1)
e^{-i\hat{H}(\tau_1 - \tau_2)}\gamma_0 \sigma(\tau_2) d \tau_2 \}&   \nonumber
\end{eqnarray}
and the Hermitian conjugated operator is
\begin{eqnarray}
\lefteqn {S^{\dagger}(\bar{\psi}_0\gamma_0, 0; \psi_t,t)=
 \exp \{\bar{\psi}_0\gamma_0  e^{i\hat{H}t} \psi_t} & &  \\
& \hspace {5mm}
 - i \int\limits^{t}_{0}[\bar{\sigma}e^{i\hat{H}(t-\tau)}\psi_t
+\bar{\psi}_0 \gamma_0 e^{i\hat{H}\tau}\gamma_0 \sigma] d\tau -
\int\limits^{t}_{0} d\tau_1 \int\limits^{t}_{\tau_1}\bar{\sigma}(\tau_1)
e^{-i\hat{H}(\tau_1 - \tau_2)}\gamma_0 \sigma(\tau_2) d\tau_2 \} & \nonumber
\end{eqnarray}

The last things we need are the integral kernels of the operator
$(-1)^{\hat F}$. They are
\bb
(-1)^F (\bar{\psi}\gamma_0, \psi) = \exp (-\bar{\psi}\gamma_0 \psi)
\ee
 Substituting the expressions (40) and (42)-(44) into the
formula (13), we get for the single mode the following evolution law:
\begin{eqnarray}
\lefteqn{\rho_{\bf p}(\bar{\psi}\gamma_0, \psi; t)=
\frac{\exp[\bar{\psi}_{\bf p} \gamma_0 \psi_{\bf p}]}
{4 e^{2 \beta \mu}(\cosh \beta\mu + \cosh \beta E)^2}\times} && \nonumber \\
 & \hspace{2mm}\exp \{ -[\bar{\psi}_{\bf p} \gamma_0 +
i \int\limits^{t}_0 \bar{\sigma}_{\bf p}e^{i\hat{H}_{\bf p}(t-\tau)}d\tau]
[1+e^{-\beta (\hat{H}_{\bf p} -\mu)}]
[\psi_{\bf p} - i \int\limits^{t}_{0}
e^{i\hat{H}_{\bf  p}(\tau-t)}\gamma_0 \sigma_{\bf p} d\tau] \}
\end{eqnarray}

One is to average it over the fluctuations $\bar{\sigma}_{\bf p},
\sigma_{\bf p}$ with the Gaussian weight
\bb
\langle \rho_{\bf p}(t)\rangle \propto
\int[d\bar{\sigma}_{\bf p}d\sigma_{\bf p}]
\, \rho_{\bf p} \,(t)
\exp(\int\limits^{t}_{0} \bar{\sigma}_{\bf p}\hat{\epsilon}^{-1}
\sigma_{\bf p} \, d\tau)
\ee
It is convenient to define the matrices of occupation numbers
\bb
\hat{n}_{\bf p} (\beta,\mu)=\frac{1}{e^{\beta(\hat{H}_{\bf p}-\mu)}+1};~~
1-\hat{n}_{\bf p} (\beta, \mu) = \bar{\hat{n}}_{\bf p}(\beta, \mu)
\ee
Note that the terms bilinear in $\sigma$ fields on the right-hand
side of Eq. (45) change the matrix $\hat{\epsilon}$ to $\hat{\epsilon}'$.
In terms of $\bar{\hat n}_{\bf p}$, the inverse of the new
operator $\hat{\epsilon}'_{\bf p}$ has the form

\begin{eqnarray}
\lefteqn{\hat{\epsilon}^{\prime}_{\bf p}=
[\hat{\epsilon}^{-1}_{\bf p} \delta(\tau_1 -  \tau_2)
- e^{-i\hat{H}_{\bf p}(\tau_1 - \tau_2)}(\bar{\hat{n}}_{\bf p})^{-1}
 \gamma_0]^{-1} =} & & \nonumber \vspace {3mm} \\
& \hspace {15mm}
\hat{\epsilon}_{\bf p}  \delta (\tau_1 - \tau_2) +
\hat{\epsilon}_{\bf p} e^{-i\hat{H}_{\bf p} \tau_1}\bar{\hat{n}}_{\bf
p}^{-1} \sum \limits_{k=0}^{\infty} [\hat{\Xi}_{\bf p}(t)
\bar{\hat{n}}_{\bf p}^{-1}]^{k}
e^{i\hat{H}_{\bf p} \tau_2}\gamma_0 \hat{\epsilon}_{\bf p} &
\end{eqnarray}
where the matrix $\hat{\Xi}_{\bf p}(t)$ is given by the expression
\bb
\hat{\Xi}_{\bf p}(t) = \int\limits^{t}_{0} d\tau
e^{i\hat{H}_{\bf p}\tau}\gamma_0
\hat{\epsilon}_{\bf p} e^{-i\hat{H}_{\bf p} \tau}
\ee
The quotient of determinants coming from the integration is
\bb
\frac{\det \hat{\epsilon}^{\prime}_{\bf p}}{\det \hat{\epsilon}_{\bf p}} =
\det [1- \bar{\hat{n}}^{-1}_{\bf p} \hat{ \Xi}_{\bf p} (t)]
\ee
One can find it either directly or by normalizing the density matrix.
After a time $t$ it becomes
\begin{eqnarray}
\lefteqn{\langle \rho_{\bf p}(\bar{\psi}\gamma_0, \psi; t)\rangle  =}
& &   \nonumber \\
 & \hspace{5mm}\det [\bar{\hat{n}}_{\bf p} - \hat{\Xi}_{\bf p}(t)]
\exp \bar{\psi}_{\bf p}\gamma_0
\{ 1-e^{-i\hat{H}_{\bf  p} t}[\bar{\hat{n}}_{\bf  p} -
\hat{\Xi}_{\bf  p}(t)]^{-1} e^{i\hat{H}_{\bf p} t} \} \psi_{\bf p} &
\end{eqnarray}

Thus the stochastic field has the following effect on the
occupation number matrix:
\bb
\hat{n}_{\bf  p}  (t)  =  e^{-i\hat{H}_{\bf p}(t-t_0)}[\hat{n}_{\bf   p}
(t_0)   +
\hat{\Xi}_{\bf p}
(t-t_0)]e^{i\hat{H}_{\bf p}(t-t_0)}
\ee
To conclude this section, we would like to stress that up to now the only
assumption used was the stochasticity of the external influence.

\subsection{Physical examples.}

Now we shall demonstrate explicitly the effect of the stochastic field on the
density matrix of fermions for the particular choice of the correlation
function. We shall consider the three-component matrices
$\hat{\epsilon}$ [see Eq. (38)]: \bb
\hat{\epsilon}_{\bf p} =\delta_{\bf p}+\chi_{\bf p}\gamma_{0}+\eta_{\bf
p}\gamma_0 \frac{\hat{H}_{\bf p}}{E_{\bf p}}
\ee
The constants $\delta, \chi$ and $\eta$ are the respective intensities.  It
will be shown that the first and third parts change the temperature and
generate fermion-antifermion pairs.  The second component acts
$C$-asymmetrically and affects the fermion number, {\em i.e.}, the difference
between the densities of particles and antiparticles.

Let us express the matrix of the occupation numbers $\hat{n}_{\bf p}$ (47) in
terms of numbers of fermions $n_{\bf p}^{+}$ and antifermions
$n_{\bf p}^{-}$, keeping apart the piece proportional to the Hamiltonian
$\hh_{\bf p}$ (37):
\bb
\hat{n}_{\bf p}=\frac{1}{2}(n^{+}_{\bf p}-n^{-}_{\bf p}+1)+\frac{1}{2}
\frac{\hat{H}_{\bf p}}{E_{\bf p}} (n^{+}_{\bf p} + n^{-}_{\bf p} - 1)
\ee
where $n_{\bf p}^{+}$ and $n_{\bf p}^{-}$ are defined by the formulae
\bb
n^{+}_{\bf p}(\mu,\beta)=\frac{1}{1+\exp \beta(E_{\bf p}-\mu)}~~~~~
n^{-}_{\bf p} (\mu,\beta)=\frac{1}{1+\exp \beta(E_{\bf p}+\mu)}
\ee
The external influence makes $\hat{n}_{\bf p}$ change according to (52).
Substituting into the definition (49) the expression (53) for
$\hat{\epsilon}$, one gets (we omit the index ${\bf p}$ up to the end of the
section).
\bb
\hat{\Xi}(t)= t(\sigma\frac{m\hat{H}}{E^2}+\chi+
\eta \frac{\hat{H}}{E})+O(\frac{\delta}{E^2})
\ee
Up to oscillations of order $\frac{\delta p}{E^{2}}$, the particle density
changes linearly with time. The first and third components do not
distinguish between particle and antiparticles, producing them in pairs.
That is characteristic of the temperature increase
\bb
\frac{n^{+}(\tau)+n^{-} (\tau)}{2} \atlim {t}{0} =
(\eta + \frac{m}{E}\delta) t
\ee
The second term is antisymmetric in fermion number
\bb
\frac{n^{+}(\tau)-n^{-}(\tau)}{2}\atlim{t}{0} =\chi t
\ee
For $\chi > 0$, one can believe it to be the external injection of
particles. Thus $\chi$ describes either the leak of particles from outside
or, depending on the sign, their drain to the bath.

Returning to conventional thermodynamic variables $\beta$ and $\mu$, we get
the following equations:
\begin{eqnarray}
\frac{1}{2} \frac{d(\mu \beta)}{dt}=\chi(1+\cosh \beta\mu \cosh\beta E)
-(\eta+\frac{m}{E}\delta)\sinh\beta\mu \sinh\beta E\\ \nonumber
\frac{1}{2} E \frac{d \beta}{dt}=\chi \sinh\beta E \sinh \beta\mu-
(\eta+\frac{m}{E}\delta)(1+\cosh \beta\mu \cosh\beta E)
\end{eqnarray}

The formulae simplify for the symmetric case $\mu = 0$
\begin{eqnarray}
\dot{\mu}=\frac{2}{\beta} \chi(1+\cosh \beta E)\\ \nonumber
\dot{\beta}=-\frac{2}{E}(\eta+\frac{m}{E}\delta)(1-\cosh \beta E)
\end{eqnarray}
We readily conclude that if initially the chemical potential was zero, then
the external fields containing only $\eta$ and $\delta$ components only
changed the temperature but did not introduce any fermion number asymmetry.

The characteristic feature of the model is the linear time variation of the
occupation numbers. We already mentioned in Section 3 that this limits the
applicability, since for physical situations $0 \leq n_{\bf p}^{+},
n_{\bf p}^{-} \leq 1$. The second limitation comes from the absence of
interaction between the modes, resulting in fully independent $\mu_{\bf p}$
and $\beta_{\bf p}$ for different values of ${\bf p}$.  Thus the uniform
heating
of the system could be modelled only by the special spectra of noise.  The
${\bf p}$-dependences of $\chi_{\bf p}, \delta_{\bf p}$ and $\eta_{\bf p}$ for
this case can be found from the equations (59) if one assumes $\mu_{\bf p}$ and
$\beta_{\bf p}$ to be the same functions of time
$\mu_{\bf p} (t) = \mu (t)$, $\beta_{\bf p} (t) = \beta (t)$
for all values of ${\bf p}$.  We shall return to
this point in the Conclusion.

\section{Conclusion}

The purpose of the present paper was to check whether stochastic classical
fermi fields are a good model for the influence of the heat bath on the
free fermi system (we'll call the situation the $F$ case).  We started from
the single fermionic degree of freedom, going on to study the evolution of
the density matrix when subject to the action of external sources.  We then
generalized the results to the Dirac fermi fields.

In the first case, the changes of the density matrix can be described by
variation of the only parameter, {\em i.e.}, the temperature.  Depending on
the sign of the noise constant, either heating or cooling of the system
takes place. The mean number of fermions depends linearly on time, with the
intensity of the noise being the scale.

The situation is richer for the Dirac fermions. Here two processes
with opposite charge parity occur.  The first is the charge-symmetric
creation of fermion-antifermionic pairs and the second is the antisymmetric
generation of (anti-)particle excess.  It could describe the particle
exchange for the charge-asymmetric environment.  In the absence of the
second component, an initially symmetric situation remains symmetric
permanently. The densities of Dirac particles change linearly with time
too.

At first sight, this situation is analogous to what happens in quantum Bose
systems interacting with stochastic Bose fields (the $B$ case). It was
shown \cite{aaa} that, although the white noise changes the density matrix,
it remains mainly (up to small corrections) the thermal one. Thus we can
talk about heating as opposed to some irregular excitation of the higher
levels.  The thermal $\rho$ matrix can be characterized by the Planckian
occupation number $n^{\rm Pl}_{\bf p}(\beta)$ which is the famous
\bb
n^{\rm Pl}_{\bf p}(\beta)=\frac{1}{\exp{\beta\omega_{\bf p}}-1}
\ee
function of temperature.  Here $\omega_{\bf p}$ is the frequency of the
oscillator in question. The noise of intensity $\lambda_{\bf p}$ changes
$n^{\rm Pl}_{\bf p}(\beta)$ linearly
\bb
n^{\rm Pl}_{\bf p}(\tau )\atlim {t}{0} \propto
\frac{\lambda _{\bf p} t}{2\omega _{\bf p}}
\ee
and that seems to be the common case.

However, the bosonic systems have some specific features. Firstly, it is
natural (see \cite{aaa}) to fix the noise constant $\lambda_{\bf p}$ to be
positive.  Thus the bosonic noise can only heat, but cannot cool. Secondly,
occupation numbers for bosonic systems are not restricted and there are no
limits on the heating.  At high temperatures, $n^{\rm Pl}(\beta \omega)
\approx (\beta \omega)^{-1}$ and the linear growth of occupation number
corresponds to the linearly increasing temperature.

Still from the point of view of possible applications, both $B$ and $F$ cases
look very similar. The main advantage to be mentioned is that here the
temperature changes much faster than in the models with dissipation.  That
could make them promising tools both for numerical simulation and for physics
of phase transitions, although today they are far from being perfect.

The drawbacks of the examples discussed stem firstly from the independence
of individual modes.  That excludes all the relaxation processes which
equilibrate temperature and chemical potential in the system.  In order to
model the uniform heating, the spectrum of the noise should be specially
adjusted to  make $\beta$ and $\mu$ the same for all degrees of freedom.
This might be a difficult problem if the system is non-linear.

The second weak spot is the \underline {reverse side} of the linear pumping
of energy into the system from outside.  There is nothing wrong about it in
the $B$-case, where it simply means that the temperature of the heat-bath
is infinite.  But fermi statistics prohibits the unlimited growth of
occupation numbers and linear laws are allowed only as the first
approximation applicable for short times.  To my mind the main
source of these inconsistencies is that
averaging over the fluctuations independently of the density matrix neglects
the thermalization.

One can think of two possible escapes from the difficulties.  The first one
is phenomenological. Let us suppose the noise intensity to be a function of
both temperature and chemical potential of the system $T, \mu$ and those of
the heat bath $T_{0}, \mu_{0}$.  If this function is zero for $T/T_{0} = \mu
/ \mu_{0} = 1$, then the stationary state will be thermodynamical
equilibrium, no matter the way in which we arrive at it.  However, the
question of the stability of this fixed point arises. It seems simpler to
treat it in the $F$ case, where signs of noise constants are not fixed.

The second way is to look for the processes balancing the modes.  The
naive belief that we shall automatically get the answer when passing from
free quantum systems to non-linear ones is deceptive.  The reason is that
introducing the non-linear terms which are usually called ``an
interaction" simply changes wave functions and shifts energy levels.  But
each of the resulting eigenstates interacts with the noise separately
and nothing stirs up the energy levels.

The quantum statistical system is different from the pure quantum one
principally because it is in  contact with the heat bath. The comparison
of the occupation numbers with the equilibrium ones takes place
permanently.  The overpopulated modes become ``noisy" and discard surplus
energy into the bath.  This scenario implies that one can naturally
include stochastic fields in realistic models of heat balance. But if
this is the case, the noise constants must be functionals of the
density matrix, and the first proof of this idea is to understand how the
overheated Bose systems could return the energy to the bath.

In conclusion, I should like to stress that, despite the
imperfections, the models discussed can still find useful applications.

\section{Acknowledgements.}

It is a great pleasure for me to thank G.W. Semenoff
whose hospitality and friendly discussions were the great stimulus for
completing this work. I'm very grateful to the community of the CERN
TH-Division
where the final version was completed. Here the graphological tolerance of
Jeanne
Rostant was not of minor importance.

\newpage

\app{The holomorphic representation for fermions.}

We shall briefly remind ourselves of some formulae, referring the reader to
\cite{ggg} for a more detailed version.  One fermionic degree of freedom is
fully
described by the occupation number, which can be either $0$ or $1$.  The wave
function is a two-component vector ${\bf f} = f^{0} |0 > + f^{1} |1>$. The
scalar product is
\bb
{\bf f g}=f^{\ast}_i g^i
\ee
The operators are $2 \times 2$ matrices acting in the following way:
\bb
( \hat{A} {\bf f})^i=A^i_j f^j
\ee
One can define the standard creation and annihilation operators
\bb
{\bf \hat{\psi}}{\bf f}=f^1 \mid 0 >~~; ~~~~~~{\bf \hat{\psi}^{+}}{\bf f}=
f^0 \mid 1 >~~; ~~~~~\{ {\bf \hat{\psi} ,\hat{\psi}^{+}} \}=1
\ee
In terms of these operators the expression (A.2) looks like
\bb
\hat{A}=A^1_1 \hat{\psi}^{+} \hat{\psi}+ A^1_0 \hat{\psi^{+}}
+A^0_1 \hat{\psi}+A^0_0 \hat{\psi}\hat{\psi}^{+}
\ee
Using the commutation rules this can be put into the normal form where the
${\bf \hat{\psi}^{+}}$ operators stand to the left of the ${\bf \hat{\psi}}$
\bb
\hat{A}=A_N ({\bf \hat{\psi}^{+}},{\bf \hat{\psi}})=(A^1_1-A_{0}^{0})
{\bf \hat{\psi}^{+}}{\bf \hat{\psi}}+A^1_0 {\bf \hat{\psi}^{+}} +A^0_1{\bf
\hat{\psi}}+A^0_0
\ee
which is the definition of the normal form of the operator $\hat{A}_{N}$.

In the holomorphic representation, one replaces vectors by binomials of
anticommuting Grassmann variables $\psi, \psi^{*}$
\bb
f^{\ast}(\psi)=f^{\ast}_0 + f^{\ast}_1 \psi;~~~
g(\psi^{\ast})=g^0+ \psi^{\ast} g^1
\ee
With the help of the corresponding rules of integration,
\bb
\int d\psi=\int d\psi^{\ast}=0;~~ \int \psi^{\ast}d \psi^{\ast}=
\int \psi d\psi=1
\ee
it is easy to see that the scalar product is given by the integral
\bb
{\bf f g}=\int d\psi^{\ast} d\psi e^{-\psi^{\ast}\psi} f^{\ast}(\psi)
g(\psi^{\ast})
\ee
The Grassmann integral operators are used instead of the operator matrices.
Their kernels are defined as
\bb
A(\psi^{\ast},\psi)=e^{\psi^{\ast} \psi}A_N(\psi^{\ast},\psi)
\ee
The action of an integral operator is defined in the following way:
\bb
\hat{A} f(\psi^{\ast})=\int d\varphi^{\ast}d\varphi e^{-\varphi^{\ast}\varphi}
A(\psi^{\ast},\varphi)f(\varphi^{\ast})
\ee

Wave functions of Fermi systems with $k$ degrees of freedom are polynomials
of $k$ Grassmann variables.  For independent states they can be factorized
into a product of binomials.  The generalization of the scalar product (A.8)
is given by
\bb
{\bf fg}=\int f^{\ast}(\psi^1,\ldots\psi^{k})g(\psi^{\ast}_1,
\ldots \psi^{\ast}_k) \prod \limits^{k}_{i=1}e^{-\psi^{\ast}_i
\psi^{i}} d\psi^{\ast}_i d\psi^{i}
\ee
The kernel of an operator which acts on the functions of $k$ variables is
given by
\bb
A(\psi^{\ast}_{1},\ldots \psi^{\ast}_k; \psi^{1},\ldots \psi^k)=
\exp(\sum^k_{i=1}\psi^{\ast}_i \psi_i)A_N(\psi^{\ast}_{1}, \ldots
\psi^{\ast}_k; \psi^{1}, \ldots \psi^{k})
\ee
and the operator acts according to the formula
\bb
{\hat A} f (\psi^{\ast}_{1},\ldots \psi^{\ast}_k)=
\int  A(\psi^{\ast}_{1},\ldots \psi^{\ast}_{k};
\varphi^{1},\ldots\varphi^{k}) f(\varphi^{\ast}_{1},\ldots \varphi^{\ast}_k)
\prod \limits^{k}_{i=1}e^{-\varphi^{\ast}_{i} \varphi^{i}}
d\varphi^{\ast}_{i} d\varphi^{i}
\ee
The wave function of a Dirac fermion is a bispinor containing four components
defining the states differing by charge and spin.  The corresponding
Grassmann polynomial should be a Lorentz scalar.  Thus the variables
$\psi^{\dagger}_{i}$, $ \psi^{i}$, $i = 1,...,4$ are bispinors as
well.\footnote{The relativistic invariance is broken in thermodynamics because
the heat bath gives the reference frame.  So we do not discuss the
difference between the true scalars $\bar{\psi} \psi$ and 0-components of
four-vectors $\psi^{\dagger} \psi$.}
The wave functions of a single fermion are given by
\bb
f(\psi^{\dagger})= f_0+\psi^{\dagger}_{i}f^{i}=f_0+\psi^{\dagger}f;~~
g^{\ast}(\psi)=g^{\ast}_{0}+g^{\dagger}\psi
\ee
This representation is handy when working with a non-relativistic
Schr\"odinger formalism.  The Lorentz-invariant Dirac expressions are
acquired after the substitution $\psi^{\dagger}\gamma_{0} \rightarrow
\bar{\psi}$.  That entails the changes of the wave function polynomials
\bb
f( \psi^{\dagger} )\rightarrow  f(\bar{\psi}\gamma_0);~~
f^{\ast}(\psi)\rightarrow f^{\ast}(\psi)
\ee
The scalar product becomes (note that $\det \gamma_{0} = 1$)
\bb
{\bf f g} =\int f^{\ast}(\bar{\psi}\gamma_0)g(\psi) e^{-\bar{\psi}\gamma_0
\psi} \prod \limits _{i=1}^4 d \bar{\psi}_i d \psi^{i}
\ee
It is quite natural that $\gamma_{0}$ appears in the operator kernels too:
$$
A(\bar{\psi}\gamma_{0}; \psi)=e^{\bar{\psi}\gamma_{0} \psi}
 A_N(\bar{\psi}\gamma_{0},\psi)
$$
Thus the operators acting between the one-fermionic states, e.g. the
Hamiltonian can be expressed in terms of the $4\times 4$ matrix $\hat{A}_N $
by the following formula:
\bb
A(\bar{\psi}\gamma_{0}; \psi)=e^{\bar{\psi}\gamma_{0} \psi}
\bar{\psi}\gamma_{0} \hat{A}_N \psi
\ee

\app{Computation of the Evolution Operator for Fermions}
\subapp{One Degree of Freedom}

The Grassmann integral kernel of the evolution operator is given by the
path integral \cite{ggg}
\bb
S(\psi^{\ast}_{t}, t;~~\psi_{0}, 0)=\int [d\psi^{\ast}d\psi]
\exp[\psi^{\ast}(t)\psi(t)+i\int \limits ^{t}_{0}(i\psi^{\ast}\dot{\psi}-
m\psi^{\ast}\psi+\psi^{\ast}{\sigma}+\sigma^{\ast}{\psi})d\tau]
\ee
The external Fermi fields are assumed to be some functions of time
$\sigma^{*}(\tau), \sigma (\tau)$ and the boundary conditions for
$\psi, \psi^{*}$ fields are
\bb
\psi(0)=\psi_0~~~~~\psi^{\ast}(t) =\psi^{\ast}_t
\ee
The Gaussian integral (B.1) is fully defined by the value of the exponent on
the classical trajectory.  The equations for the latter are the following:
\bb
\begin{array}{rcccr}
i\partial_{\tau}\psi & - & m\psi & = & -\sigma \\
-i \partial_{\tau}\psi^{\ast} & - & m\psi^{\ast} & = & -\sigma^{\ast}
\end{array}
\ee
where $\partial_{\tau} = \frac{\partial}{\partial \tau}$.
The solutions compatible with the boundary conditions (B.2) are
\bb
\begin{array}{lcccr}
\psi(\tau) & = & e^{-im\tau}\psi_{0} & + & i\int \limits^{\tau}_{0}
e^{-im(\tau-\xi)} \sigma(\xi)d\xi \\
\psi^{\ast}(\tau) & = & \psi^{\ast}_{t}e^{im(\tau-t)} & + &   i\int \limits
^t_{\tau} e^{im(\tau-\xi)}\sigma^{\ast}(\xi)d\xi
\end{array}
\ee

Substituting them into the definition (B.1), we obtain for the kernel of the
evolution operator
\begin{eqnarray}
\lefteqn{S(\psi^{\ast}_{t}, t; \psi_{0}, 0)=
\exp[\psi^{\ast}_{t}e^{-imt}\psi_{0}}  & & \\  & \hspace{5mm}
+i\int\limits^{t}_{0}(\sigma^{\ast}
e^{-im\tau}\psi_{0}+\psi^{\ast}_{t} e^{-im(t-\tau)}\sigma)d\tau-
\int\limits^{t}_{0} d\tau_{1}\int\limits^{\tau_1}_{0}d\tau_2 \sigma^{\ast}
(\tau_1)e^{-im(\tau_1 -\tau_2)}\sigma(\tau_2)] & \nonumber
\end{eqnarray}
and the Hermitean conjugated expression is
\begin{eqnarray}
\lefteqn{S^{\dagger}(\psi^{\ast}_{0}, 0; \psi_{t}, t)=
\exp[\psi^{\ast}_{0}e^{imt}\psi_{t}}  & & \\  & \hspace{5mm}
-i\int\limits^{t}_{0}(\sigma^{\ast}
e^{im(t-\tau)}\psi_{t}+\psi^{\ast}_{0} e^{im\tau}\sigma)d\tau-
\int\limits^{t}_{0} d\tau_{1}\int\limits^{t}_{\tau_1}d\tau_2 \sigma^{\ast}
(\tau_1)e^{-im(\tau_1 -\tau_2)}\sigma(\tau_2)] & \nonumber
\end{eqnarray}

\subapp{The Dirac Fermions}

In the case of Dirac fermions, the evolution operator has the kernel \cite{ggg}
\begin{eqnarray}
\lefteqn{S(\bar{\psi}_{t}\gamma_{0}, t;\psi_{0}, 0)=} & & \\ & \hspace{5mm}
\int[d \bar{\psi}d\psi]\exp \{\bar{\psi}(t)\gamma_{0} \psi(t)
+i \int\limits ^{t}_{0} [ \bar{\psi} (\hat{p}-m)\psi+
\bar{\sigma}\psi+\bar{\psi}\sigma ] d \tau \} & \nonumber
\end{eqnarray}
Here $\bar{\psi}, \psi, \bar{\sigma}, \sigma$ are bispinors. As before, the
fields $\bar{\sigma} (\tau), \sigma (\tau)$ are functions of time, and the
boundary conditions for the $\bar{\psi}, \psi$ fields are
\bb
\psi(0)=\psi_{0}; ~~~ \bar{\psi}(t)=\bar{\psi}_{t}
\ee
The equations of classical trajectories now look like
($\vec{\partial}_{\mu} \psi = - \psi \stackrel{\leftarrow}{\partial}_{\mu} =
\partial_{\mu} \psi$)
\bb
\begin{array}{ccccc}
\hfill ~(\hat{p}-m) \psi &  = & \hfill
{}~(i\gamma_{\mu}\stackrel{\rightarrow}{\partial_\mu}-m) \psi & = &
-\sigma \\
\bar{\psi} (\hat{p}-m)~\hfill & = & \bar{\psi}
(i\stackrel{\leftarrow}{\partial_\mu}\gamma_{\mu }-m)~\hfill  & = &
-\bar{\sigma}
\end{array}
\ee
It is convenient to use the plane wave expansion to solve them:
\bb
\psi(\tau, x)=\sum_{\bf p} \psi_{\bf p}(\tau)e^{-ip_{i}x^{i}},~~
\bar{\psi}(\tau, x)=\sum_{\bf p} \bar{\psi}_{\bf p}(\tau)e^{ip_{i}x^{i}}
\ee
The boundary conditions for $\bar{\psi}_{\bf p}, \psi_{\bf p}$ are given by
the corresponding Fourier components $\bar{\psi}_{i},\psi_{f}$.
The bispinors  $\bar{\psi}_{\bf p}, \psi_{\bf p}$ are functions of time only
and satisfy the differential equations
\bb
\begin{array}{ccc}
\hfill ~(i\gamma_{0}\stackrel{\rightarrow}{\partial_\tau}-\gamma_{i}p_{i}-m)
\psi_{\bf p} & = &
-\sigma_{\bf p}  \\
\bar{\psi_{\bf p}} (i\stackrel{\leftarrow}{\partial_\tau}\gamma_{0}-
p_{i} \gamma_{i}-m) ~\hfill & = &
-\bar{\sigma}_{\bf p}
\end{array}
\ee
The solutions compatible with the boundary conditions (B.10) are
\bb
\begin{array}{rclcl}
\psi_{\bf p}(\tau) & = & e^{-i\hat{H}_{\bf p}\tau}\psi_{\bf p}(0) & + & i
\int\limits^{\tau}_{0} e^{-i\hat{H}_{\bf
p}(\tau-\xi)}\gamma_{0}\sigma_{\bf p}d\xi \\
\bar{\psi}_{\bf p}(\tau) & = &
\bar{\psi}_{\bf p}(t)\gamma_{0}e^{i\hat{H}_{\bf p}(\tau-t)}\gamma_{0} & + &
i\int\limits ^{t}_{\tau}d\xi \bar{\sigma}_{\bf p}
e^{i\hat{H}_{\bf p}(\tau-\xi)}\gamma_{0} \end{array}
\ee
Substituting them into the path integral (B.6) gives
\begin{eqnarray}
\lefteqn{S(\bar{\psi}_{t}\gamma_{0}, t;
\psi_{0},0)=\exp\{\bar{\psi}_{t}\gamma_{0} e^{-i\hat{H}t}\psi_{0} + } & & \\ &
\hspace{5mm}\nonumber i \int\limits^{t}_{0}[\bar{\sigma}
e^{-i\hat{H}\tau}\psi_{0}+\bar{\psi}_{t}\gamma_{0}e^{-i\hat{H}(t-\tau)}
\gamma_{0}\sigma]d\tau -
 \int\limits ^{t}_{0}d\tau_{1}
\int\limits ^{\tau_{1}}_{0}\bar{\sigma}(\tau_{1})
e^{-i\hat{H}(\tau_{1}-\tau_{2})}\gamma_{0}\sigma(\tau_{2})d\tau_{2}\}&
\end{eqnarray}
and the Hermitian conjugated operator is (we recall that $\psi^{\dagger} =
\bar{\psi}\gamma_{0}; \; \bar{\gamma}^{\dagger}_{0} = \gamma_{0}, \;
\gamma^{\dagger} = - \gamma_{i}$)
\begin{eqnarray}
\lefteqn{S^{\dagger}(\bar{\psi}_{0}\gamma_{0}, 0;
\psi_{t},t)=\exp\{\bar{\psi}_{0}\gamma_{0} e^{+i\hat{H}t}\psi_{t}+ } & & \\ &
\hspace{5mm}\nonumber i \int\limits^{t}_{0}[\bar{\sigma}
e^{i\hat{H}(t-\tau})\psi_{t}+\bar{\psi}_{0}\gamma_{0}e^{i\hat{H}\tau}
\gamma_{0}\sigma]d\tau -
 \int\limits ^{t}_{0}d\tau_{1}
\int\limits ^{t}_{\tau_{1}}\bar{\sigma}(\tau_{1})
e^{-i\hat{H}(\tau_{1}-\tau_{2})}\gamma_{0}\sigma(\tau_{2})d\tau_{2}\}&
\end{eqnarray}
In the last two formulae (B.12, B.13), we have summed up the components and
returned from the ${\bf p}$- to the ${\bf x}$-representation.  We have
therefore
omitted the index ${\bf p}$.  However it can be reinserted since as long as
the theory is free these formulae are true for each of the Fourier harmonics
as well.

\end{document}